\documentclass[referee,floatfix]{aa} 
\usepackage{graphicx}

\begin{document}

     \title{Constraining Galactic Models Through Parallax and Astrometry
      of Microlensing Events}


\author{Sohrab Rahvar
        \inst{1,2} \fnmsep \thanks{\email{rahvar@sharif.edu}}
        \and
        Sima Ghassemi
        \inst{1} \fnmsep
          \thanks{\email{sima\_gh@mehr.sharif.ac.ir}}
     }

\offprints{S. Rahvar}

 \institute{
 Department of Physics, Sharif University of
Technology, P.O. Box 11365-9161, Tehran, Iran
 \and
 Institute for
Studies in Theoretical Physics and Mathematics, P.O. Box
19395-5531, Tehran, Iran
 }

\date{Received *****; Accepted *****}

\abstract{In recent years various models for the Galactic
distribution of massive compact halo objects (MACHOs) have been
proposed for the interpretation of microlensing toward the Large
Magellanic Cloud (LMC). A direct way to fit the best model is by
measuring the lens parameters, which can be obtained by measuring
the Einstein crossing time and the parallax effect on the
microlensing light curve and by astrometry of centroids of images.
In this work, the theoretical distribution of these parameters is
obtained for the various power-law Galactic dark halo models and
MACHO mass functions (MF). For self-lensing as one of the models
for the interpretation of LMC events, the maximum shift of image
centroids and the parallax parameter are one order of magnitude
smaller than for models with dark halos. This can be used as a
test for the self-lensing, although the shifting of image
centroids $0.02$ {\it mas}, for the LMC events is unlikely to be
observed by the astrometric missions such as FAME, GAIA and SIM.
 }
\authorrunning{S. Rahvar and S. Ghassemi}
\titlerunning{Identification of Galactic Model}
\maketitle

\section{Introduction}
One of the primary candidates for the dark matter in the Galaxy is
the massive compact halo object (MACHOs) that can be in baryonic
or even non-baryonic form. Paczy\'nski (1986) proposed the
gravitational microlensing method as an indirect means of
detecting MACHOs. The microlensing effect is the magnification of
a background star while a MACHO passes the line of sight of an
observer to a background object. Due to the small separation
between the images produced by the lensing, it is impossible to
resolve them by ground-based telescopes and only the magnification
and its duration can be measured by observing the
light curve of the background star.\\
For one decade many groups contributed in this experiment and
observed stars in the Large and Small Magellanic Clouds to measure
the contribution of MACHOs to the dark halo mass and estimate the
mass of MACHOs themselves. Due to the degeneracy between lens
parameters (i.e. mass of the lens, MACHO distance and the
transverse velocity), it is impossible to obtain them uniquely
just by measuring the duration and magnification of events. The
degeneracy problem also renders us unable to find the distribution
of MACHOs in the Galaxy although one can learn much from
statistical methods by analyzing the duration distribution of
events. Furthermore, the low number of events observed in the
direction of LMC causes considerable Poisson fluctuation; to
properly constrain Galactic models we will need many more events
to be detected as will be carried out by the next generation of
microlensing surveys (\cite{bec04}). One of the alternative ways
to overcome the problem of statistical fluctuation is improving
the quality of light curves, measuring the other observable
parameters which can break the degeneracy
between the lens parameters. \\
Apart from the duration of microlensing events, there are two more
observable parameters that can be measured by accurate observation
of light curves. Gould (1992) showed that the effect of the
rotation of earth around the Sun on the microlensing light curve
is a slight deviation of the light curve from symmetry. The
possibility of observing this (parallax) effect has been studied
towards the Galactic bulge (\cite{buc97}) and Large Magellanic
Cloud (\cite{gou98}). An observational strategy to have
sufficiently accurate microlensing light curves (with a
significant parallax signature) has been proposed to distinguish
two extreme models of the Milky Way and to test the hypothesis of
self-lensing (Rahvar et al. 2003). However, using the parallax
effect is not presently sufficient to distinguish between Galactic
models with the small number of microlensing
events.\\
The displacement of the image centroids due to microlensing can
also be measured with high resolution astrometry. The typical
angular separations of multiple images formed by microlensing in
the direction of Magellanic Clouds is $1$ milli arcsecond ({\it
mas}) (\cite{pac96,pac98}) and the image centroid is expected to
move about $ 0.1$ {\it mas}.
The astrometric missions such as
FAME\footnote{http://www.usno.navy.mil/FAME/},
SIM\footnote{http://planetquest.jpl.nasa.gov/SIM/sim\_index.htm}
and GAIA\footnote{http://astro.estec.esa.nl/GAIA/} which will be
launched in 2000s, achieve this high precision. Astrometric
accuracies anticipated for those missions are: $50$
micro arcsecond ({\it $\mu$as}) for FAME, and $10$ {\it $\mu$as}
(or possibly higher) for SIM and GAIA, where this accuracy depends
on the apparent magnitude of the stars. SIM can do astrometry for
stars as faint as 20th magnitude and it can measure the shift of
image centroids of most of the LMC stars (\cite{raj01}). This
positional shift of images
can also be used to break one more degree of degeneracy. The
advantage of astrometric microlensing is that the probability of
observing a microlensing event is larger than that of photometric
microlensing (\cite{mir96,dom00,hon01}).\\
In this work we obtain the distributions of event duration, and
two parallax and astrometry parameters for various power law dark
halo models with various MACHO mass functions (MF) and also LMC
models (so-called self-lensing). We show that for the self-lensing
model, the parallax and the astrometric
parameters are one order of magnitude smaller than for typical dark halo lensing.\\
The organization of the paper is as follows: in Section 2 we
discuss the degeneracy problem in microlensing and show how
astrometry accompanying the parallax measurements can break the
degeneracy between the lens parameters. In Section 3 we introduce
power law dark halo models in the Milky Way and obtain the
theoretical distributions of event duration, parallax and
astrometry parameters for various MACHO mass distributions. In
section 4 we discuss the hypothesis of self-lensing and the
distribution of observable parameters in this model. The results
are discussed in Section 5.
\section{Non-Standard microlensing and degeneracy breaking}
The Einstein crossing time of the microlensing events $(t_e)$,
degenerately depends on the mass, distance and the transverse
velocity of the lenses. To break this degeneracy, we can measure
both the deviation of the light curve from the Paczy\'nski form
due to the parallax effect and the displacement of image centroid
during the lensing. In this section
we give a brief account on these effects.\\
\subsection{Parallax effect on the microlensing light curve}
As the Earth is rotating around the Sun, its velocity component,
projected on the deflector plane, is not negligible compared to
the transverse speed of the deflector. The apparent velocity of
the deflector with respect to the line of sight is a cycloid
instead of a straight line and this effect causes the light curve
of the microlensing event to deviate from that of a Heliocentric
observer (\cite{alock95}).
The deviation of the light curve from simple microlensing can be
described by the parameter $\delta u$ which is the projection of
the Earth orbital radius ($a_{\oplus}$) in the deflector plane in
units of the Einstein radius and the projected transverse speed of
the deflector ($\tilde{v}$) on the Earth's orbit. $\delta u$ and
$\tilde{v}$ are defined as follows:
\begin{eqnarray}\label{1}
\delta u &=& a_{\oplus}(1-x)/R_E,\\
\tilde{v}&=& \frac{R_E}{t_E(1-x)},
\label{2}
\end{eqnarray}
where $R_E$ is the Einstein radius and $x$ is the ratio of the
observer-lens distance to the observer-source distance. Equations
(1,2) show that by measuring the signature of the parallax effect
in the microlensing light curves, one can constrain the mass and
the distance of the lenses. This effect can reduce one degree of
degeneracy between the lens parameters.
\subsection{Astrometric microlensing}
The angular separations of the images due to the microlensing of
the background stars at the LMC, lensed by the dark halo MACHOs,
are of order of one {\it mas}. The image centroid sweeps out
nearly one {\it mas} during the microlensing event. Not only the
ground-based telescopes, due to the seeing limit, nor the HST, due
to the diffraction limit, can achieve the resolution of {\it mas}
shifts in image centroid, however observations based on
interferometry in space may be able to do so. The shift of image
centroid during the lensing can be derived as follows:
\begin{equation}\label{2}
\overrightarrow{\delta\theta}=\frac{1}{2+u^{2}}\overrightarrow{u}\theta_{E},
\end{equation}
where $u$ is the separation between lens and the source in the
lens plane normalized to the Einstein radius, $\theta_{E}$ is the
angular size of the Einstein radius
($\theta_{E}=\frac{R_{E}}{D_{ol}}$, where $D_{ol}$ is the
observer-lens distance). For $u=\sqrt{2}$ the angular shift
obtains its maximum value
$\delta\theta_{max}=\frac{\theta_{E}}{\sqrt 2}$. Measuring
$\delta\theta_{max}$ yields the angular Einstein radius and 
constrains the mass and distance of the lens from the observer.
\subsection{Degeneracy breaking}
The Einstein crossing time, $t_{E}$, depends on three parameters:
the mass, $M$, transverse velocity, $v_t$, and distance of lens
from the observer, $D_s$, through
\begin{equation}
t_E = \frac{1}{v_t}\sqrt{\frac{4GMD_s}{c^2}x(1-x)}.
\end{equation}
On the other hand the parallax and astrometry parameters ($\delta
u$ and $\delta\theta_{max}$) give two more equations depending on
the mass and the distance of the lens. Using these three
equations, we can obtain the lens parameters as following:
\begin{eqnarray}
\label{9} x&=&\frac{a_{\oplus}}{a_{\oplus}+\sqrt{2}D_{s}\delta u
\delta\theta_{max}},\\
\label{10} M
&=&\frac{1}{2\sqrt{2}}\left(\frac{c^{2}a_{\oplus}\delta\theta_{max}}{G\delta
u}\right), \\
\label{11}
v_{t}&=&\frac{\sqrt{2}}{t_{e}}\left(\frac{a_{\oplus}\delta\theta_{max}D_s}{a_{\oplus}+\sqrt{2}D_{s}\delta
u \delta\theta_{max}}\right).
\end{eqnarray}
\section{Constraining Galactic Models via Degeneracy Breaking}
In the last section, we showed that measuring the duration of
events along with the astrometric and parallax parameters can
break the degeneracy between the lens parameters. In spite of the
large optical depth of astrometric microlensing compared to the
photometric one (\cite{hon01}), we need to do both astrometric and
photometric microlensing to get the mass and the distance of the
lenses. The mass and distance of lenses can be obtained by
measuring the two parameters $\delta u$ and $\delta\theta_{max}$
where their dependence on the distance of lenses from the observer
is as $\sqrt{\frac{1-x}{x}}$. This means that measuring the
distance of lenses is more accurate the closer the lens is. The
fraction of localized lenses among the overall microlensing events
is expressed by the observation efficiency, which is a function of
the accuracy of the apparatus and the observational strategy. The
expected distributions of $\delta u$ and $\delta\theta_{max}$ can
be obtained by convolving the theoretical distributions with the
observational efficiency of the experiment.\\
Here we obtain the theoretical distributions of $t_E$, $\delta u$
and $\delta\theta_{max}$ in various Galactic models for
microlensing events in the direction of the LMC. The relevant
components of the Galaxy, in the LMC microlensing events, are the
Galactic disk, halo and the LMC itself. For the dark halo, we have
considered a large set of axisymmetric models, so-called power-law
models (\cite{eva94}). To model the density of the thin and thick
disks of the Milky Way, we use a double exponential function
(\cite{bin87}). Eight power-law halo models with a corresponding
disk can be combined to build various Galactic models: these are
the Standard Model (S), Medium halo (A), Large halo (B), Small
halo (C), Elliptical (D), Maximal disk (E) and two thick disk
models (F \& G) where we follow the designations of Alcock et al. (1996).\\
One of the elements for generating $t_E$, $\delta u$ and
$\delta\theta_{max}$ is the mass function (MF) of MACHOs. The
tradition has been to use a single MACHO mass (i.e a Dirac-Delta
function). Here we use two Dirac-Delta and spatially varying MFs.
Kerins and Evans (1998) and recently Rahvar (2005) have shown that
a spatially varying MF for the MACHOs, not only can fit the
distribution of microlensing event durations in the direction of
LMC, but can also resolve the contradiction between the
microlensing results and other astrophysical observations such as
the unseen white dwarfs in the Galactic halo
(\cite{opp01,tor02,spa04}). The Dirac-Delta MF is given by
$\delta(M-M_{macho})$, where $M_{macho}$ in the various Galactic
models is given in Alcock et al. (1996, 2000). The spatially
varying MF has the form $MF(r) = \delta(M-M(r))$ where $M(r)$ scales as
$M_{U}\left(\frac{M_{L}}{M_{U}}\right)^{r/R_{halo}}$, where $M_U$
and $M_L$ are the upper and lower limits of mass of MACHOs, $r$ is
the distance from the center of Galaxy and $R_{halo}$ is a halo
scale parameter. The mean mass for the spatially varying MF in the
standard dark halo model from the likelihood analysis is about the
brown dwarf mass (\cite{rah05}). The physical meaning of this type
of spatially varying MF is that the inner part of dark halo
contains heavier MACHOs than the outer part. The optimized
parameters for the spatially varying MF of MACHOs in power-law
halo models, to be compatible with the LMC microlensing candidates
of MACHO experiment is given in Rahvar (2005). The MF of the
Galactic disk also can be taken from the results of observations
with the
{\it Hubble Space Telescope} (Gould, Bahcall \& Flynn 1997). \\
The distribution of event duration in the various Galactic models can
be obtained by numerical simulation (Alcock et al. 1996). To
obtain the distributions of $\delta u$ and $\delta\theta_{max}$
during an observation time of $T_{obs}$, we perform a Monte-Carlo
simulation, using the probability function of lenses along our
line of sight.
The rate of microlensing events due to
the MACHOs reside between the distances $x$ and $x+dx$ from the observer
is given by:
$$\frac{d\Gamma}{dx} =
4\sqrt{\frac{G D_{s}}{M(x)c^2}x(1-x)}v_t\rho(x), $$ where
$M(x)$ can be substituted from the desired MF. The location of the
selected lenses is used to calculate corresponding $\delta u$ and
$\delta\theta_{max}$. Figures 1 \& 2 show the results for the
distributions of event duration, parallax parameter and maximum
centroid shift in the power-law halo models. For the models with
dominant halo such as $S, A, B, C$ and $D$, the distribution of
$\delta u$ and $\delta\theta_{max}$ looks similar, while they are
different from the disk dominant models ($E, F$ and $G$). The
distribution of $\delta\theta_{max}$ is sensitive to the type of
MFs. For the spatially varying MF, the distribution of
$\delta\theta_{max}$ shifts to significantly
smaller values than single MACHO mass MFs.
\begin{figure}
\begin{center}
\includegraphics[height=16.2cm,width=14.5cm]{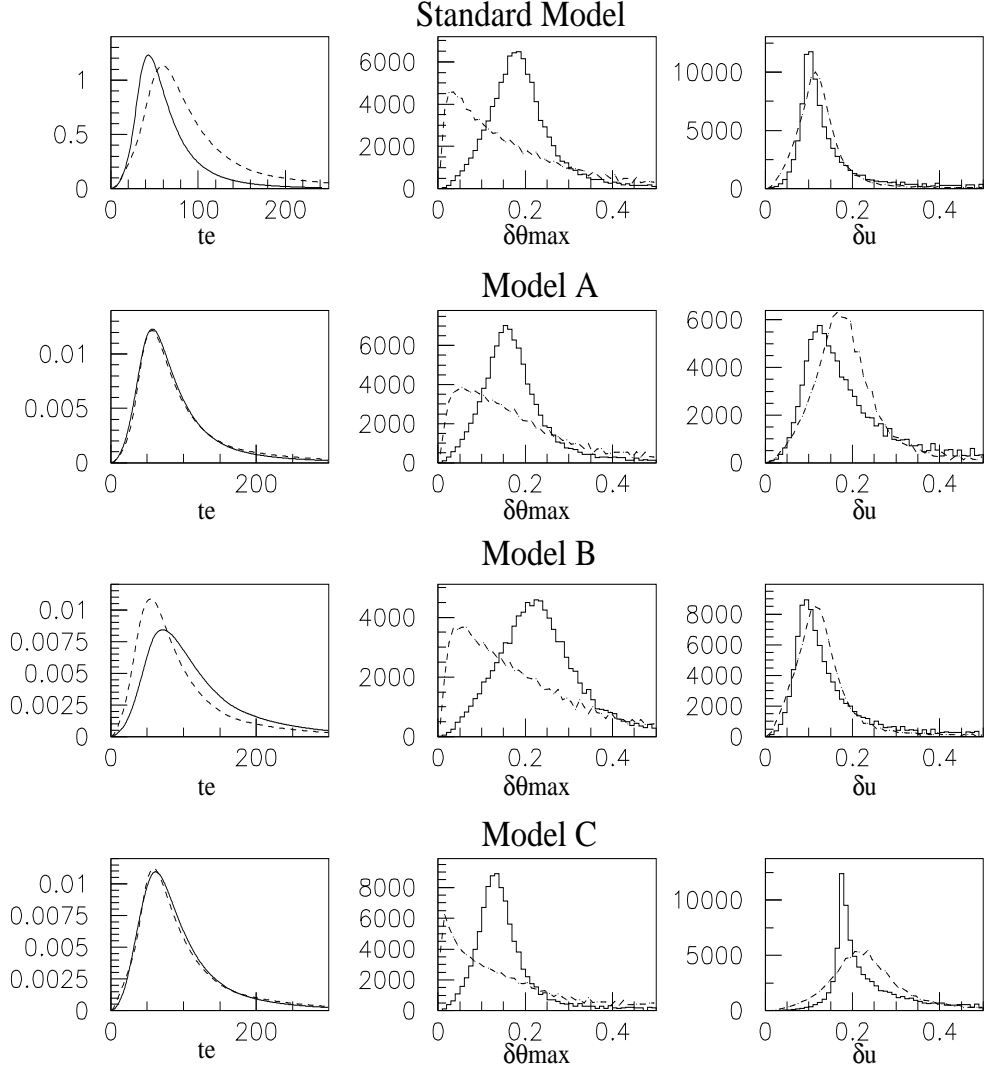}
\caption{\label{l} Distributions of event duration, $t_e$, (in
days, total number of events is normalized to one), parallax parameter, $\delta u$, and
astrometry parameter, $\delta \theta_{max}$, (in {\it mas}) for
the Galactic halo models of S, A, B \& C. The solid curves are the
distributions derived using the Dirac-delta MF and the dashed
lines for a spatially varying MF.}
\end{center}
\end{figure}
\begin{figure}
\begin{center}
\includegraphics[height=16.2cm,width=14.5cm]{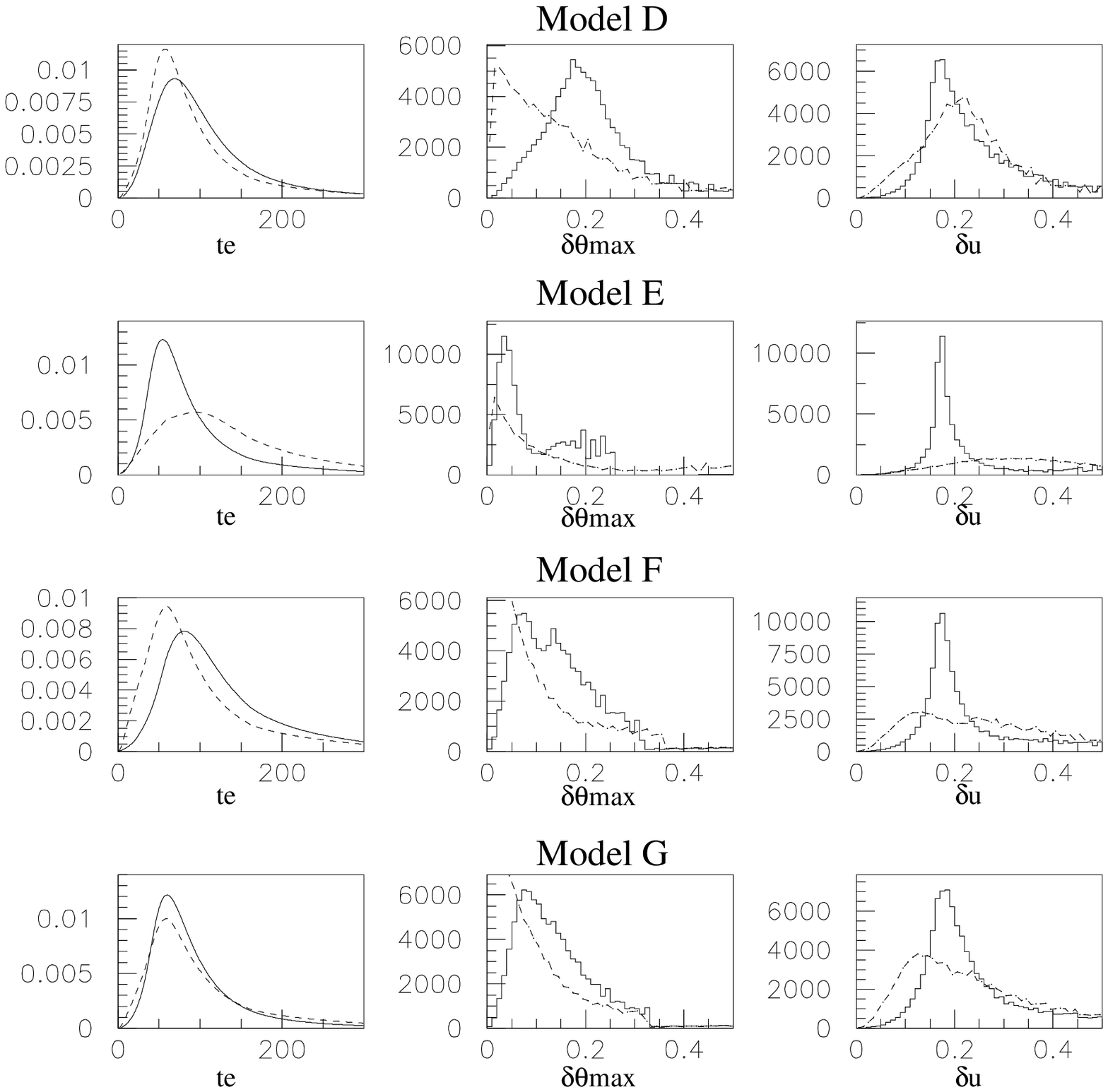}
\caption{ \label{2} Distributions of event duration, $t_e$, (in
days, total number of events is normalized to one), parallax parameter, $\delta u$, and
astrometry parameter, $\delta \theta_{max}$, (in {\it mas}) for
the Galactic halo models of D, E, F \& G. The solid curves are the
distributions derived using the Dirac-delta MF and the dashed
lines for a spatially varying MF. }
\end{center}
\end{figure}
\section{Testing the hypothesis of LMC self-lensing}
\begin{figure}
\begin{center}
\includegraphics[height=16.2cm,width=14.5cm]{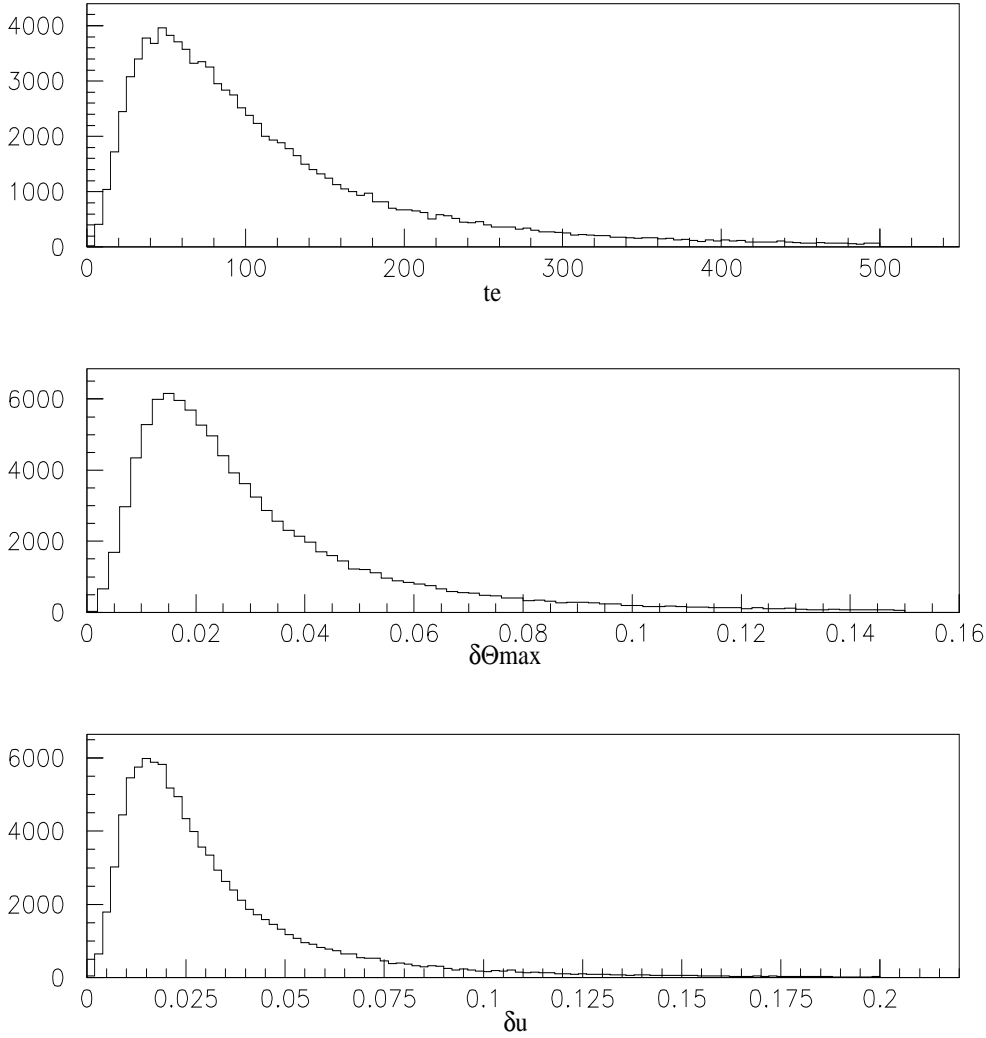}
\caption{\label{3} The distributions of duration of events $t_e$
(in days), parallax parameter $\delta u$ and the astrometry
parameter $\delta \theta_{max}$ (in {\it mas}), for the
self-lensing events. The typical values of $\delta u$ and $\delta
\theta_{max}$ for the self-lensing events comparing with Figs. (1
\& 2) are one order of magnitude smaller than for the halo .}
\end{center}
\end{figure}

Due to the low optical depth of microlensing events in the
direction of the Magellanic Clouds, the EROS and MACHO experiments
have obtained only 3 and 13-17 microlensing candidates,
respectively (\cite{las00,alc00,mil05}). These studies lead to an
upper limit of about 3\% (in EROS experiment) to 20\% (in MACHO
experiment) obtained for the contribution of MACHOs to the dark
halo and a mean MACHO mass of about that of a white dwarf
(\cite{las00,alc00,mil05}). However, the expected white dwarfs
have not been seen by studies such as proper motion measurements
(\cite{opp01,tor02,spa04}). This inconsistency can be resolved by
(i) considering a spatially varying MF for the MACHOs in the
Galactic halo or (ii) considering the LMC stars playing the role
of lenses. If most of MACHOs reside in the Galactic halo, the
timescale analysis of LMC microlensing candidates implies that the
typical mass of lenses is of the order $0.5 M_{\odot}$ while for
the SMC events, the mass of lenses is estimated to be 2 to 3
$M_{\odot}$ (\cite{sah03}). This
inconsistency can be solved by considering the lenses as belonging to the LMC.\\
The other motivation for the possibility of self-lensing was
presented by Sahu (2003), where the number of localized binary
lenses was compared with the overall microlensing candidates. If
half of the MACHOs are considered to be in a binary system
(similar to the stars), then we expect to observe half of the
microlensing events via binary lenses. On the other hand only $20
\%$ of binary lenses are expected to show the caustic crossing
feature, then we should see the caustic crossing for $10\%$ of
lenses. Two of the Magellanic Clouds' microlensing candidates had
caustic a crossing feature and by breaking the degeneracy between
the lens parameters, they are localized within the Magellanic
Clouds. According to this rough estimation, about $20$
microlensing events are expected via LMC self-lensing; this number
is about the same number of events as
have been observed. \\
Using the parallax effect and astrometric centroid shift
observations can test the hypothesis of self-lensing. We used a
disk model for the LMC (\cite{gyu00}) and obtained the
distributions of $t_e$, $\delta u$ and $\delta\theta_{max}$ for
self-lensing. Comparing Figure (\ref{3}) with Figures (\ref{1})
and (\ref{2}) shows that for self-lensing events $\delta u$ and
$\delta\theta_{max}$ are about one order of magnitude smaller than
that of halo lensing. For the halo lensing the mean value of the
maximum shift for the centroid of images is about $0.2$ {\it mas}
while for the self-lensing, it is about $0.02$ {\it mas}.
Comparing the resolution of the astrometric missions with the
centroid shift for the halo and LMC lenses, shows that FAME, SIM
and GAIA have enough resolution to observe the halo lensing while
for the self-lensing events the resolution is of the same order as
the centroid shift and it is unlikely to the detectable.

\section{Conclusion}
In this work we study the degeneracy breaking between
the lens parameters by measuring the duration of events and
deviation of light curves due to the parallax effect and
the maximum displacement of image centroids during the
lensing.\\
We used various Galactic models and obtained
the theoretical distributions of observable parameters. In
order to have the observed distribution of parameters we
need to convolve these distributions with details of the
observing program such as the photometric and
astrometric accuracy and the sampling rate. One of the
applications of astrometry with the parallax measurements is
testing the hypothesis of self-lensing. The maximum shift of the
centroid of images and parallax parameter for the halo lensing are
one order of magnitude large than the self-lensing ones.\\
In practice, observations to measure the three parameters of
microlensing can be done by a ground-based telescope accompanying
a space-based interferometry telescope. The ground-based telescope
alerts one to the microlensing events and undertakes photometric
measurements and the space-based telescope measures the
displacement of image centroids. Rahvar et al.(2003) proposed a
strategy for observation of microlensing events based on both
alert and follow-up telescopes. They obtained the observational
efficiency for the detection of parallax effect and estimated the
number of microlensing events essential to distinguish between two
extreme Galactic models and to test the self-lensing hypothesis.
This type of observational strategy
can be extended by adding a space-based astrometric telescope.\\

\begin{acknowledgements}
The authors thank the anonymous referee giving useful comments and
improving the text of paper.
\end{acknowledgements}

\end{document}